\documentclass[prc,showpacs,preprintnumbers,amsmath,amssymb]{revtex4}
\usepackage{graphicx}

\begin{document}

\title{Tri-axial deformation in nuclei with realistic NN interactions}
\author{E. N. E. van Dalen and H. M\"uther}
\affiliation{Institut f\"ur Theoretische Physik, \\
Universit\"at T\"ubingen, D-72076 T\"ubingen, Germany}

\begin{abstract}
The structure of finite nuclei is investigated by employing an interaction
model  which is based on the low-momentum interaction $V_{lowk}$. It is 
supplemented by a density-dependent contact interaction fitted to reproduce 
the saturation properties of infinite nuclear matter within the Hartree-Fock 
approach. The calculations of finite nuclei are performed in a basis of 
plane waves discretized in a cartesian box of appropriate size.
As a first example the structure of Ne isotopes is considered ranging from $^{18}$Ne 
to the neutron drip line. Rather good agreement is obtained for the bulk properties 
of these nuclei without any free parameter. The basis is also appropriate to 
describe other deformed nuclei and the transition from discrete nuclei to 
homogeneous matter which is supposed to occur in the crust of neutron stars.
\end{abstract}


\maketitle

\section{Introduction}
The availability of radioactive ion beam facilities in several laboratories - like SPIRAL2 at Ganil and the future GSI facility
FAIR in Germany - has made it possible to measure properties of many neutron-rich nuclei far away from the valley of $\beta$-stability.
The study of these exotic nuclear systems is an interesting field of physics. It is
driven by the expectation of observing  nuclear properties, which are
different from those encountered so far, i.e. near the valley of stability, such as  special deformations and neutron halos.

The role of deformations has been considered to some extent. Some attention has been devoted to the consequences for the shell closures far from
stability. Deformations  can influence  the double $\beta$-decay rate. However, almost all calculations of the double $\beta$-decay matrix elements performed so far
have been made assuming spherical symmetric nuclei, whereas we know that in many cases, e.g. $^{76}$Ge, they are not. A study of the double $\beta$-decay of $^{48}$Ti reaches the conclusion that a large mismatch of deformation between the parent
and the grand-daughter nuclei can reduce the value of nuclear matrix elements by factors as large as 2-3~\cite{caurier:2008}.
Therefore, an accurate description of the geometrical structure of nuclei is important, in particular for neutron-rich nuclei far away from the line of  $\beta$-stability. The geometrical structures of these nuclei can be quite complex compared to deeply-bound, double magic nuclei such as $^4$He, $^{16}$O, and  $^{40}$Ca. Furthermore, neutron-rich deformed nuclei can be regarded as the starting point of the transition
to infinite homogeneous matter, which should occur in the outer crust of a neutron star. 

Most of the studies on nuclear system far away from the valley of stability have been performed using density functional theories (DFT).
Non-relativistic\cite{bender:2003} as well as covariant\cite{niksic:2011} density functional theories have been developed during the last twenty years and successfully been applied to describe a large variety of nuclear phenomena with high precision. The basis
of this success, however, are the careful parameterizations of the underlying density functionals. The variety of developments and the
number of different parameterizations is growing so rapidly that only a few specialists are able to keep an overview in specific
subfields of DFT.

The alternative would be ab initio calculations, which use a realistic model of the nucleon-nucleon (NN) interaction, based on the description of the NN
scattering data. However, there are at least four obstacles on the way to derive properties of finite nuclei in such an ab initio approach. 

The first problem arises from the fact that also the realistic NN interaction is not unique, the number of models, which all fit the NN scattering data is large, ranging from rather stiff models for local NN interactions\cite{arg1} to softer 
One-Boson-Exchange models\cite{mach1} or the chiral models\cite{chirar,chira1,chira2}, which have received a lot of attention, recently. 
The second obstacle is due to the fact that even the softer version of realistic NN interaction models contain strong central and tensor
components, which make it inevitable to employ many-body theories well beyond the mean-field approximation (see e.g. \cite{polls00} and references cited there). However, obstacle number three, even employing sophisticated many-body theories one is in general not able to reproduce the saturation point of nuclear matter, the simplest nuclear structure one can imagine. Finally, the effective NN interactions resulting from such a microscopic approach are rather complicate and are typically determined in terms of matrix elements using an appropriate basis of single-particle states. So problem number four is the choice of an appropriate set of basis states to describe weakly bound nuclei allowing for a general deformation.

In the following we will address these obstacles for a nuclear structure calculation a little bit more in detail and present an approach, which is a kind of compromise: A nuclear structure calculation based on the realistic NN interaction supplemented by a DFT
correction with a minimal number of adjustable parameters. First applications of this approach in nuclear matter and deeply bound
nuclei have been presented in \cite{vandalen:2009b}. It is the aim of the work presented here, to test this approach in the description
of weakly bound nuclei.
  
A possible solution of the problems one and two of an ab initio approach, outlined above, is the use of a low-momentum interaction, which can be derived
from a realistic NN interaction by means of renormalization
techniques~\cite{bogner:2001,bogner:2005,bogner:2007,bozek:2006,vandalen:2010}. It is obtained by separating the low-momentum and high-momentum components of a realistic NN interaction.
This means that all short-range components of the NN
interaction are integrated out. Therefore, a non-perturbative
treatment of NN correlations is not necessary anymore. 
If the cut-off $\Lambda$ is appropriately chosen, i.e. around $\Lambda$ = 2 fm$^{-1}$, the resulting low-momentum
interaction $V_{lowk}$ turns out to be independent of the underlying realistic
interaction. So using $V_{lowk}$ we have a unique NN interaction, which describes the NN scattering data up to the pion threshold
and can be used in a mean field approach.

However, using the low-momentum interaction $V_{lowk}$ a new problem arises. Hartree Fock calculations employing $V_{lowk}$ in infinite nuclear matter produce 
a binding energy per nucleon increasing with density in a monotonic way~\cite{bozek:2006,kuckei:2003}. Hence,
the emergence of a saturation point is prevented, the problem number three mentioned even gets worse. While in Brueckner-Hartree-Fock calculations using realistic NN interactions the calculated saturation point does not agree with the empirical data, one does not even get
one using $V_{lowk}$. 

A way out of this problem is to add three- or many-nucleon forces in the Hamiltonian. There are various possible sources for such
many-nucleon interactions. The renormalization procedure, which in the case of the two-nucleon problem leads to the $V_{lowk}$ interaction, 
gives rise to many-nucleon forces when applied to systems with three or more nucleons. Such effective three-nucleon forces, which arise from the renormalization with respect to high momentum components within a pure nucleonic model, should provide similar effects as e.g. the density and energy dependence of the Brueckner G-matrix and therefore support the saturation mechanism. However, also sub-nucleonic
degrees of freedom or relativistic effects may be represented in terms of a medium dependent effective NN force or three-nucleon forces. 

In the present study we do not care about the origin of such many-nucleon forces but just accept the fact that a Hamiltonian based
on $V_{lowk}$ must be supplemented to be used in calculations of nuclear structure. This supplement has been chosen to contain basic  contact interaction terms of a non-relativistic DFT like the Skyrme model\cite{sk1,bv81} and adjust the parameters to reproduce the empirical saturation
point of symmetric nuclear matter. This has been done in  \cite{vandalen:2009b} and there it could also be demonstrated that this approach yields very reasonable results for the bulk properties of closed shell nuclei like $^{16}$O and $^{40}$Ca.

These double magic nuclei can be described with a spherical geometry using spherical coordinates~\cite{coraggio:2005,coraggio:2006}. However, non-spherical nuclei require a geometry allowing for the description of more complex structures. Therefore, some recent investigations~\cite{martini:2011,ebran:2011}  applied cylindrical coordinates to describe axially-symmetric-deformed nuclei. Although 
one can describe simple prolate and oblate deformations with the cylindrical coordinates, the description of more complex tri-axial structures is not so easy. Therefore, we have chosen a cartesian geometry to allow the description of tri-axial geometrical structures in these nuclei. 

Furthermore, nuclear structure calculations of finite nuclei with realistic forces employ a set of appropriate single-particle wave functions as a set of basis states,
typically those of a harmonic oscillator (HO) potential. This choice of the HO basis seems to be quite reasonable if one wants to describe the
structure of deeply-bound closed-shell nuclei. However, the HO basis may be questionable when weakly-bound nuclei
close to the proton or neutron drip line are to be considered, since it is not appropriate to describe the tail  of the single-particle
wave functions for the weakly-bound valence nucleons. For this kind of studies a basis of plane wave (PW) states may be more suitable~\cite{vandalen:2009b}.

One aim of this study is to establish a tool, which allows a microscopic description of tri-axial geometrical structures in nuclei and in the transition from
isolated nuclei to homogeneous matter based on a realistic NN interaction. Therefore, we will apply a cubic Wigner-Seitz cell using cartesian coordinates
and a tri-axial PW basis. This geometry will allow us to extend our investigation to tri-axial geometrical structures
of nuclei. The nuclear structure calculations will be performed in the framework of a Hartree-Fock (HF) calculation using a
$V_{lowk}$ potential supplemented with a contact interaction. 

As an example we apply this approach to investigate the density distributions of neon isotopes. An issue of interest is the situation of N = 20 shell closure in neon, which lies far away from the valley of 
$\beta$-stability. Compared to recent investigations~\cite{martini:2011,ebran:2011}, our calculations are performed using a tri-axial PW basis  and cartesian coordinates in a cubic Wigner-Seitz cell. This geometry will allow us to extend our investigation to tri-axial geometrical structures instead of only axially-symmetric deformations.
Furthermore, we use a realistic interaction, a $V_{lowk}$ potential, instead of a phenomenological interaction. In addition, we will go beyond the description of stable neon isotopes to the situation with unbound neutrons, which will have some relevance for the transition from discrete nuclei to homogeneous matter which is
supposed to occur, e.g. in the crust of neutron stars.

The plan of this paper is as follows. The method is treated in Sec.~\ref{sec:method}. The procedure to renormalize the low-momentum
interaction and obtain the matrix elements of $V_{lowk}$ is reviewed in
Sec.~\ref{sec:Vlowk_mat_el}. In this section we also describe our interaction model
supplementing $V_{lowk}$ by an appropriate contact interaction. In Sec.~\ref{sec:HF_Hamil}, we work out
how to solve the HF equation in a basis of tri-axial plane wave
states in cartesian coordinates.
The results for density distributions of the neon isotopes are presented and discussed in
Sec.~\ref{sec:R&D}. Finally, Sec.~\ref{sec:SC} contains a summary and the
conclusions of our work.

\section{Method}
\label{sec:method}

\subsection{Model of the NN interaction}
\label{sec:Vlowk_mat_el}

The main idea of $V_{lowk}$ interaction is to separate the long-range part from the short-range part  of a realistic NN interaction
and restrict the nuclear
structure calculation to the low-momentum components, i.e. the long-range part.
This long-range or low-momentum part of the
NN interaction is fairly well described in terms of meson-exchange, whereas
the quark degrees of freedom are important in the description of the
short-range or high-momentum components of the NN interaction.
In other words, a model space, which accounts for the low momentum
degrees of freedom, is defined and the effective
Hamiltonian is  renormalized  for this low-momentum regime to account for
the effects of the high-momentum components, which are integrated out.
In our case, the matrix elements for $V_{lowk}$ are determined by using the unitary-model-operator approach (UMOA),
a model space technique to disentangle low  and high momentum  components.
This approach has been described in the literature~\cite{suzuki:1982}. Therefore, we will restrict ourselves purely to the
basics.

First, we define the projection operator $P$, which projects onto the low-momentum
subspace. In our case, i.e. the effective two-nucleon problem, the operator $P$ projects onto the model space of two-nucleon wave functions with
relative momenta $k$ smaller than a chosen cut-off $\Lambda$. Next, the projection operator $Q$ is defined. The  projection operator $Q$ projects onto the complement of this subspace, i.e. the 
high-momentum subspace. In the UMOA approach, the aim is to define a unitary transformation $U$ in
such a way that the transformed Hamiltonian does not couple $P$ and $Q$ spaces. The technique to obtain this unitary transformation has
been described in Ref.~\cite{bozek:2006,fuji:2004}. With the help of this unitary transformation, the low momentum interaction can be obtained.
It is defined as
\begin{equation}
  V_{eff} = U^{-1}(h_0 + v_{12}) U - h_0,
\end{equation}
where $v_{12}$ represents the bare NN interaction and the starting Hamiltonian $h_0$ denotes the one-body part of the two-body system.
The phase shifts obtained from solving the Lipmann-Schwinger equation for NN scattering using this $V_{eff}$ with a cut-off
$\Lambda$ are identical to those obtained from solving the Lipmann-Schwinger equation  for the original realistic interaction
$v_{12}$ without a cut-off. If the original interaction  is
realistic, this means that this $V_{eff}$ will also be a realistic interaction and reproduces the NN scattering phase shifts up to a cut-off
$\Lambda$. This $V_{eff}$ will from now on be called $V_{lowk}$.

An interesting feature is that the resulting $V_{lowk}$ is found to be essentially
model independent. In other words, this $V_{lowk}$  potential will be independent of the
underlying realistic interaction $v_{12}$, if the cut-off $\Lambda$ is chosen around
$\Lambda$ = 2 fm$^{-1}$. In the calculations discussed in section \ref{sec:R&D} below, we actually
consider the proton-neutron part
of the CD Bonn potential~\cite{mach1} for the bare interaction $v_{12}$ using a cut-off parameter $\Lambda$ = 2 to obtain $V_{lowk}$.

The fact that the $V_{lowk}$ does not induce any short-range correlations into the nuclear wave function,
seems to lead to  problems in the description of the
saturation behavior of nuclear matter. Due to this lack of short-range
correlation effects, the emergence of a saturation point is prevented in calculations of symmetric nuclear
matter~\cite{bogner:2005,kuckei:2003}. Hence, three-body interaction terms 
or density-dependent two-nucleon interactions are
needed to obtain saturation in isospin symmetric nuclear matter.
Therefore, the effective interaction $V_{lowk}$ is supplemented by a 
contact interaction, which we previously presented in Ref.~\cite{vandalen:2009b}. This contact interaction has the form
\begin{equation}
\Delta\mathcal{V} = \Delta\mathcal{V}_0 + \Delta\mathcal{V}_3,
\label{eq:contact}
\end{equation}
with
\begin{equation}
\Delta\mathcal{V}_0 = \frac{1}{4}t_0[(2+x_0)\rho^2-(2x_0+1)(\rho_n^2+\rho_p^2)]
\end{equation}
and
\begin{equation}
\Delta\mathcal{V}_3 = \frac{1}{24}t_3 \rho^{\alpha}[(2+x_3)\rho^2-(2x_3+1)(\rho_n^2+\rho_p^2)],
\end{equation}
where $\rho_p$ and $\rho_n$  refer to the local proton and neutron densities. Furthermore,  the matter density is denoted as $\rho=\rho_p+\rho_n$.
The parameters of the contact interaction are obtained in the following way: 
the parameters $\alpha=0.5$ and $x_0=0.0$ have been fixed. The three remaining parameters $t_0$,
$t_3$, and $x_3$
have then been fitted to reproduce the empirical saturation point of nuclear
matter and the symmetry energy at saturation density. The results for these fitting parameters are listed in
Table~\ref{table:CT_Param}.
\begin{table}
\begin{center}
\begin{tabular}{|c|c|c|c|}
\hline
\ \ $t_0$ \ [MeV fm$^3$] \ \  & \ \ $t_3$
\ [MeV fm$^{3+3\alpha}$] \ \ & \ \ \ \ \ \ \ \
$x_3$ \ \ \ \ \ \ \ \ \\
\hline
584.1  & 8330.7 & -0.5   \\
\hline
\end{tabular}
\end{center}
\caption{\label{table:CT_Param}
The fitted parameters $t_0$, $t_3$ and $x_3$ defining the contact interaction of
Eq.~(\protect{\ref{eq:contact}}).}
\end{table}

\subsection{Hartree-Fock Calculations in a  tri-axial Plane Wave Basis}
\label{sec:HF_Hamil}
While the Hartree-Fock or mean-field contribution arising from the contact interaction 
can easily be evaluated in configuration space, the effective interaction $V_{lowk}$ is
non-local and is typically defined in terms of antisymmetrized two-body matrixelements
\begin{equation}
< \alpha \gamma\vert V_{lowk} \vert \beta \delta> \,,\label{eq:vlow2}
\end{equation} 
using a set of single-particle basis states $\vert \alpha >,\,\vert \beta > \dots $. This set of
basis states should be
appropriate to describe the self-consistent single-particle states emerging from the Hartree-Fock
calculation for deformed nuclei, but also allow the evaluation of the matrix elements of $V_{lowk}$
in a straightforward and simple manor. 

As we are interested in the description of nuclei with weakly bound valence nucleons and close
to the neutron drip line, we did not use a basis of harmonic oscillator states but considered
a set of plane-wave states with periodic boundary conditions on the surface of a cubic box with 
a box size of $L$ in each cartesian direction, i.e. a volume of $L^3$. This means that
the cartesian components of the momenta are discretized and we obtain e.g. for the $x$-direction
$$
k_{xj} = \frac{2\pi\,j}{L}\quad \mbox{for}\quad j=0,1,\dots
$$
and corresponding values for the $y$ and $z$ direction. Therefore each basis state $\alpha$ is
identified by the 3 cartesian components forming the momentum $\vec k$ and the quantum number
$s=\pm 1/2$ defining the projection of the nucleon spin in the directions of the z-axis.
Therefore the orthonormal single-particle basis states can be written in configuration space
\begin{equation}
\Phi_\alpha (\vec r) = \frac{1}{\sqrt{L^3}} \exp(i\vec k\cdot\vec r) \,\chi_s\,,\label{eq:basis}
\end{equation}
with $\chi_s$ denoting the two-component Pauli spinor. The Hartree-Fock single-particle states
are then expanded in this basis
\begin{equation}
\vert \Psi_n > = \sum_{\alpha=1}^N \vert \alpha > < \alpha \vert\Psi_n > =
\sum_{\alpha=1}^N
c_{n\alpha} \vert \alpha >\,,\label{eq:hfexpan}
\end{equation}
and the $V_{lowk}$ part of the Hartree-Fock Hamiltonian can be written in
terms of the two-body matrix elements, of (\ref{eq:vlow2})
\begin{equation}
< \alpha \vert H_{lowk}^{HF} \vert \beta > = \sum_{\gamma,\delta} < \alpha \gamma\vert
V_{lowk} \vert \beta \delta> \rho_{\gamma\delta}\,\label{eq:defhlowk},
\end{equation}
where $\rho_{\gamma\delta}$ represents the single-particle density matrix
\begin{equation}
\rho_{\gamma\delta} = \sum_{n=1}^A c_{n\gamma}^* c_{n\delta} = \sum_{n=1}^A<\Psi_n\vert \gamma>< \delta \vert\Psi_n >\,,\label{eq:1dens}
\end{equation}
with a sum including all occupied HF states $\vert\Psi_n>$. It is straightforward to supplement this $V_{lowk}$
part of the Hartree-Fock Hamiltonian by the corresponding terms for the kinetic energy, the contact interaction
of (\ref{eq:contact}) including the so-called rearrangement terms and the Coulomb potential for the protons
\cite{pgreinh,Bonche:2005,Ryssens:2014}
\begin{equation}
< \alpha \vert H^{HF} \vert \beta > = <\alpha\vert H_{kin} + H_{lowk}^{HF} + \Delta\mathcal{V}^{HF} +
V_{Coul}^{HF}\vert\beta>\,.\label{eq:hfham}
\end{equation}
This HF Hamiltonian is diagonalized
\begin{equation}
\sum_{\beta=1}^N < \alpha \vert H^{HF} \vert \beta > c_{n\beta} = \varepsilon_n c_{n\alpha}\,,\label{eq:hfeq}
\end{equation}
to obtain the expansion coefficients $c_{n\alpha}$ and the HF single-particle energies $\varepsilon_n$. Since
the definition of HF Hamiltonian depends on the expansion coefficients (see e.g. eq.(\ref{eq:defhlowk})) the
evaluation of $H^{HF}$ and its diagonalization must be iterated in the usual way until a self-consistent
solution is obtained. We initiate this iteration by approximating the HF states by appropriate eigenstates of
a tri-axial deformed Woods Saxon potential.
We note that the Coulomb contribution in (\ref{eq:hfham}) has been calculated using the so-called Slater approximation for the
exchange part\cite{pgreinh,Ryssens:2014} and that the energy for the spurious center of mass motion has been subtracted from the energy after the variational calculation. 

We still have to specify some details concerning the definition and the truncation of the PW basis defined in
(\ref{eq:basis}) and used in (\ref{eq:hfexpan}). Most of the calculations discussed below have been performed
assuming for the size of the cartesian box a length $L$ of 18 fm. It turned out that stable results were obtained if the expansion in (\ref{eq:hfexpan}) was restricted to momenta $\vec k$ with $\vert\vec k\vert \leq$
2 fm$^{-1}$, which leads to a typical number of 739 different momenta $\vec k$ to be included. Including the spin degree of freedom this implies for the number of basis states $N=1478$ in (\ref{eq:hfexpan}) for neutron states and the same number of coefficients representing the proton states. The only symmetry, which has been imposed, was to assume the HF states to be states of good parity. This symmetry ensures that the center of the
density distribution is in the center of the cartesian box.  

For the description of spherical nuclei it is of course easier to use a spherical basis like e.g. the eigenstates of a free particle moving in a spherical cavity with radius $R$ as it has been done e.g. in
\cite{vandalen:2009b}. In this case the wave functions of the orthogonal basis can be separated 
in a radial part and an angular part
\begin{equation}
\Phi_{iljm}(\vec r) =  R_{il}(r)
{\cal Y}_{ljm} (\vartheta,\varphi)\,.\label{eq:sbasis}
\end{equation}
The angular part  ${\cal Y}_{ljm}$ represent the spherical harmonics including
the spin degrees of freedom by coupling the orbital angular momentum $l$ with
the spin to a single-particle angular momentum $j$. For the case of spherical nuclei one obtains
single-particle states with definite angular momentum $j$ and a degeneracy of $2j+1$. The expansion
of the HF states is restricted to an expansion of the radial wave function and typically it is sufficient
to consider of the order of 10 expansion coefficients. 

The spherical basis of (\ref{eq:sbasis}) can also be used to describe deformed nuclei. In this case one must
allow the HF states to include admixtures of different angular momenta $ljm$ and would obtain again larger
dimensions for the expansion of the HF states.  A disadvantage of the spherical basis is the fact that it does
not really allow for a good description of the limit of homogeneous nuclear matter. This is obvious if one uses a basis of radial functions (i.e. spherical Bessel functions) with a node at the border of the cavity $r=R$.
A Slater determinant built from such basis states for free particles leads to a non-uniform density distribution. But also using a basis with mixed boundary conditions for different partial waves does not lead
to a homogeneous density distribution.

The plane wave basis of (\ref{eq:basis}) constrained to a cartesian box contains the limit of homogeneous matter 
in a trivial way. In fact, this limit has been used as one check of the computer code and we were able
to reproduce results of infinite matter. The HF calculation of $^{16}$O and $^{40}$Ca starting from a spherical initial wave function has been another test. In this case the HF iteration preserves the symmetry of the
initial state and also the calculation in the cartesian basis yield spherical solutions, which are within the
numerical accuracy identical to the solutions obtained in the spherical basis (\ref{eq:sbasis}).

\section{Results and discussion}
\label{sec:R&D}
The model for the NN interaction discussed above has been developed in \cite{vandalen:2009b} and used
to describe the properties of closed shell nuclei $^{16}$O and $^{40}$Ca. It is the aim of the investigations presented here to test this approach for the evaluation of properties of open shell nuclei. Special attention will be paid to the determination of the neutron drip line and the formation of non-spherical density distributions. As an example we consider the structure of the Ne isotopes with even nucleon numbers $A$ ranging from  $A=18$ to $A=32$. We also consider the case of $A=34$, which is beyond the neutron drip line and an example of $A=42$ with 32 neutrons, which should partly be evaporated to the continuum.

\begin{table}
\begin{tabular}{|c | c | c |}
\hline
\ \ isotope \ \ & \ \ cal. E/A [MeV] \ \ & \ \ exp. E/A  [MeV] \ \ \\
        
\hline
$^{18}$Ne & -7.20  &  -7.34 \\
$^{20}$Ne & -7.96  &  -8.03 \\
$^{22}$Ne & -8.03  &  -8.08 \\
$^{24}$Ne & -8.02  &  -7.99 \\
$^{26}$Ne & -7.63  &  -7.75 \\
$^{28}$Ne & -7.31  &  -7.39 \\
$^{30}$Ne & -7.01  &  -7.07 \\
$^{32}$Ne & -6.72  &  -6.67 \\
$^{34}$Ne & -6.17  &   -    \\
    \hline
    \end{tabular}
\caption{Results for the energies per nucleon
using the $V_{low-k}$ plus contact
term interaction model in the tri-axial plane wave basis. The
calculated energies have been corrected by the one-body center-of-mass energy. The experimental
values are taken from Ref.~\cite{audi:1993}.}
\label{table:E_Neon}
\end{table}

The results for the binding energy per nucleon are displayed in table \ref{table:E_Neon} and compared
to the experimental data taken from Ref.~\cite{audi:1993}. The difference between the calculated values and
the experimental data is for all cases below 3 MeV for the total binding energy and therefore comparable to
the deviations typically obtained in pure phenomenological DFT calculations\cite{ebran:2011}. This is very encouraging noting that the interaction model considered here is based on a realistic NN interaction supplemented by a phenomenological part with only three parameters adjusted in the description of nuclear matter.

Also the position of the neutron drip-line is reproduced, as one can see from a comparison of the total binding energies deduced from this table. The total binding energy evaluated for $^{34}$Ne (34 times 6.17 MeV) is smaller than the corresponding binding energy of $^{32}$Ne. 

\begin{figure}[!h]
\begin{center}
\includegraphics[width=0.6\textwidth] {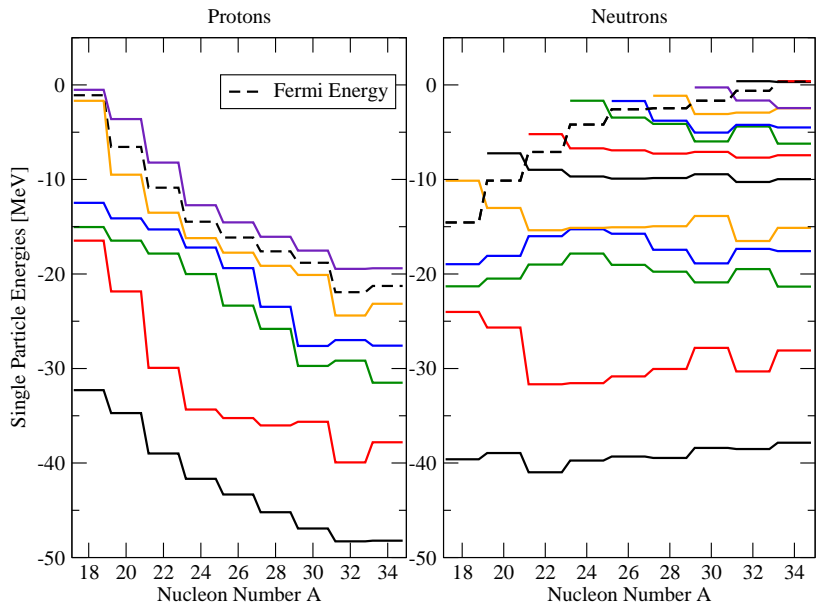}
\caption{(Color online) Single-particle energies for protons (left panel) and neutrons (right panel) obtained from Hartree-Fock calculations for various Ne isotopes with even nucleon number $A$. Note that each line represents the position of two degenerate states. The energies of all occupied and the lowest unoccupied level
are presented. The dashed lines represent the position of the Fermi energy.}
\label{fig:singpart}
\end{center}
\end{figure}

The position of the drip-line can also be extracted from the spectra of the single-particle states displayed
in the two panels of Fig.~\ref{fig:singpart} for protons and neutrons, respectively. The HF calculations have been performed without a constraint to time reversal symmetry. Nevertheless, the resulting single-particle energies form pairs of almost degenerate states exhibiting differences up to a few hundred keV. The plot shows the position of the lower of such two single-particle energies.

The single-particle energies for the protons decrease in a rather monotonic way with increasing nucleon number
$A$. This reflects the strong attraction of the proton-neutron interaction leading to a stronger binding for
the protons with an increasing number of the neutrons. 

The single-particle energies for the neutrons, displayed in the right panel of Fig.~\ref{fig:singpart} do not show such a monotonic behavior. Most of the single-particle energies change only very little with neutron number. This is true in particular for $A\geq 22$, i.e. the isotopes with more neutrons than the $^{22}$Ne, which is the isotope with largest binding energy per nucleon. 

Going from $A$ to $A+2$ only the position of the states, which are not occupied in the isotope with $A$ nucleons but
get occupied in the $A+2$ isotope receive a stronger binding. Inspecting the neutron single-particle energies, which are close to the Fermi energy in the cases of $A$=32 and $A$=34, the position of the neutron drip-line discussed above gets confirmed. While all neutron states occupied in $^{32}$Ne have single-particle energies
below 0, one pair of states to be occupied in $^{34}$Ne has an energy larger than 0. These states are almost degenerate with the lowest particle states above the Fermi energy.

\begin{figure}[!h]
\begin{center}
\includegraphics[width=0.6\textwidth] {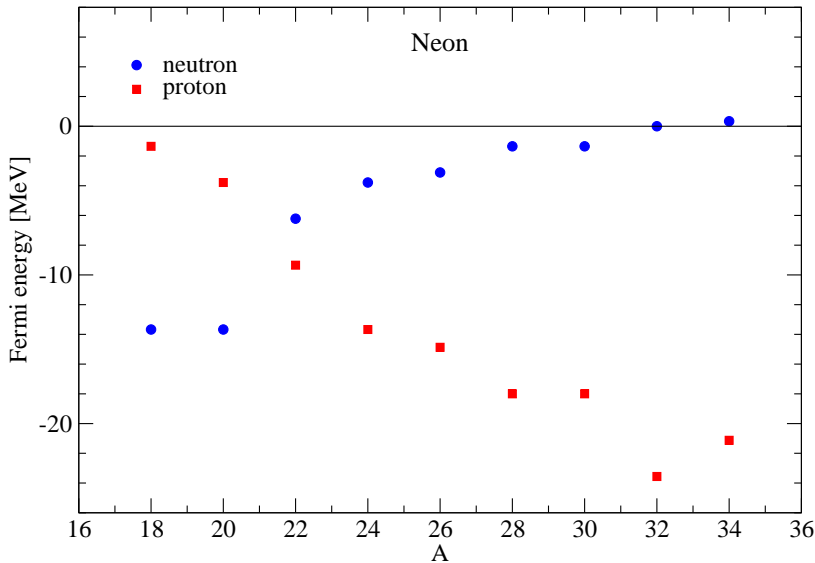}
\caption{(Color online) The Fermi energies for the neutrons and the protons in neon isotopes.}
\label{fig:EFNe}
\end{center}
\end{figure}

Some of these features are also represented by the Fermi energies, which we define as the arithmetic mean value of the highest occupied state and the lowest unoccupied state and show in Fig.~\ref{fig:EFNe}.
% .  

The Fermi energy for the protons steadily decrease with increasing nucleon number, while the Fermi energy for the neutrons increase very weakly for $A\geq$22 and cross the zero at $^{34}$Ne.

\begin{figure}[!h]
\begin{center}
\includegraphics[width=0.6\textwidth]{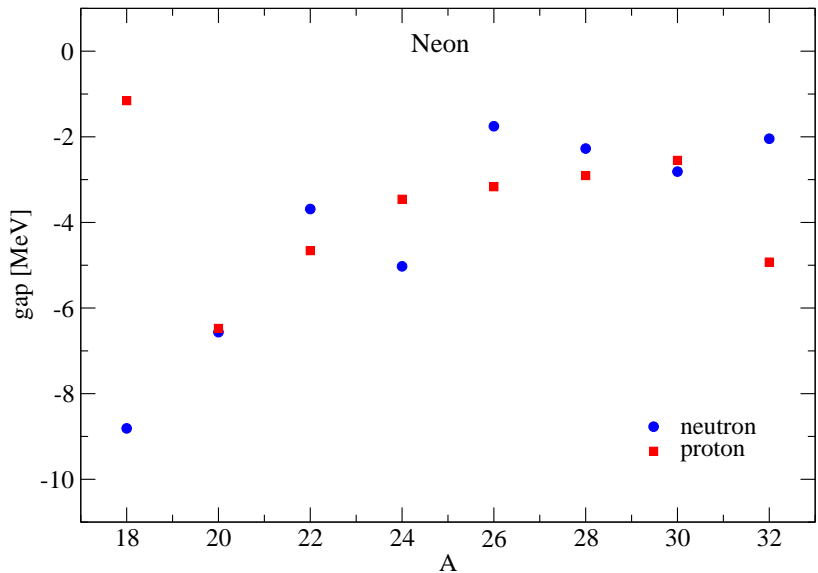}
\caption{(Color online) The energy gap between the last occupied state and first unoccupied state for various neon isotopes.}
\label{fig:gapneon}
\end{center}
\end{figure}

Another interesting quantity to be deduced from the single-particle spectra is the energy gap, i.e. the energy
difference between the highest occupied state and the lowest unoccupied state. The corresponding results are shown in Fig.~\ref{fig:gapneon}. The absolute values for the gaps in the single-particle spectra for protons and neutrons are large than 6 MeV in the case of $^{20}$Ne. For all other isotopes considered at least one of those values for the gap are smaller than 4 MeV, which is an indication that for these nuclei $T=1$ pairing correlations may be relevant.
 
The saturation point of nuclear matter is identified by the energy and the corresponding value for the baryon density. A many-body calculation is considered successful only, if both quantities are reproduced. In close analogy we should be satisfied with a microscopic theory describing the bulk properties of finite nuclei, if the binding energies of the nuclei as well as the radii of the charge distribution are described. 

\begin{figure}[!h]
\begin{center}
\includegraphics[width=0.6\textwidth] {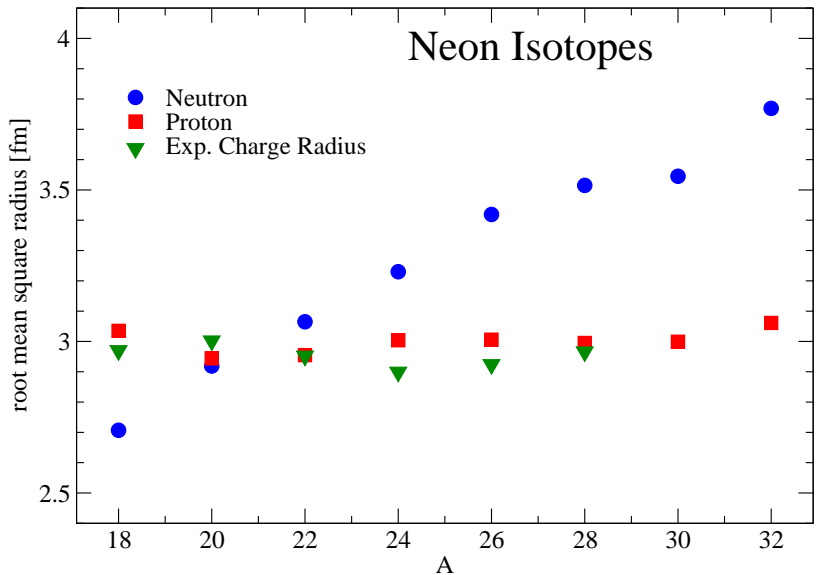}
\caption{(Color online) The neutron and proton root mean square radius of neon isotopes. The experimental data for the charge radii have been obtained from measurements of the optical isotope shifts in \cite{marinova}}
\label{fig:rms}
\end{center}
\end{figure}

Corresponding results for the are displayed in Fig.~\ref{fig:rms}. The agreement between the radii calculated for the proton distribution and the values for the charge distributions\cite{marinova} are remarkable. This demonstrates that the contact interaction of (\ref{eq:contact}) is appropriate to stabilize the density of infinite nuclear matter as well as
the surface of light nuclei at the empirical value.

Fig.~\ref{fig:rms} also displays the root mean square radii calculated for the density distributions for the neutrons. The radius increases with the number of neutrons, representing the development of neuron skin for $N>Z$ with a rather monotonic slope. A small deviation from this monotonic slope is observed only at $A$=30, which corresponds to a neutron number $N=20$ a value at which the $sd$-shell in a simple shell-model would be filled. 
 
\begin{figure}[!h]
\begin{center}
\includegraphics[width=0.8\textwidth] {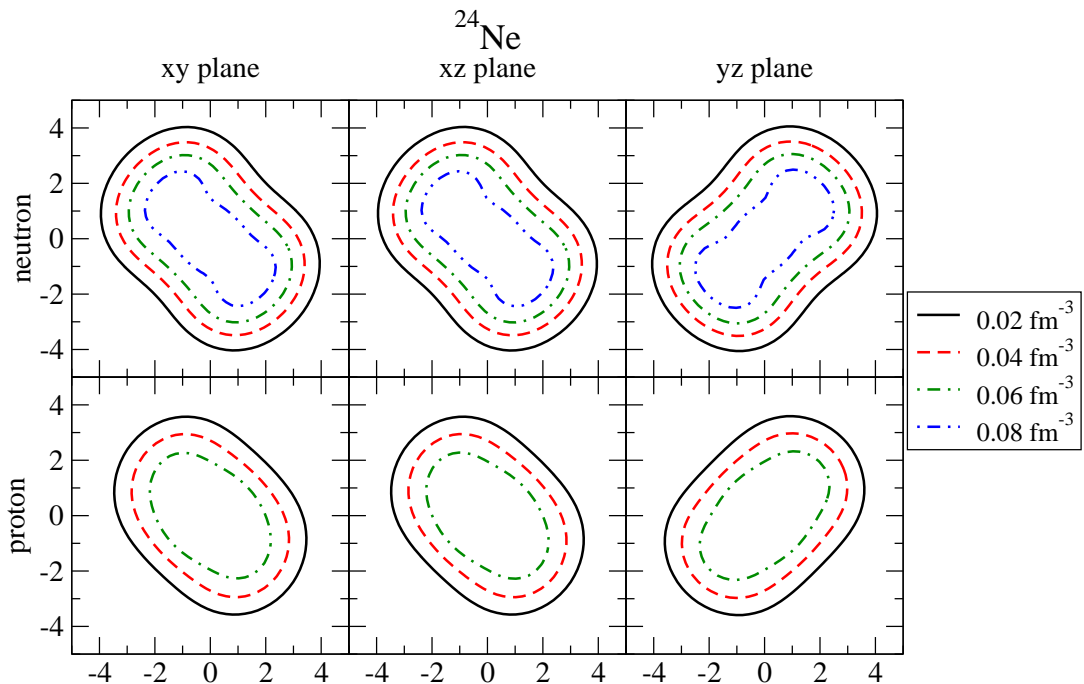}
\caption{(Color online) Contour plots of the neutron and proton density distributions of $^{24}$Ne in the xy, xz, and yz plane.}
\label{fig:Ne24}
\end{center}
\end{figure}

The radius, however, is only one observable to be deduced from the density distribution. 
In Fig.~\ref{fig:Ne24}, the neutron and proton density distributions of $^{24}$Ne are shown in terms of
contour plots projected to the three cartesian planes.    

These  plots show that the nucleus is elongated in one direction, which is characterized by a vector pointing to the corner of the cartesian box, in which the calculation is performed, with positive values for the $y$ and $z$ coordinate and negative value for the $x$ coordinate. The elongation axes for protons and neutrons are identical, which reflects the attraction between protons and neutrons. 

\begin{figure}[!h]
\begin{center}
\hbox{
\includegraphics[width=0.45\textwidth]{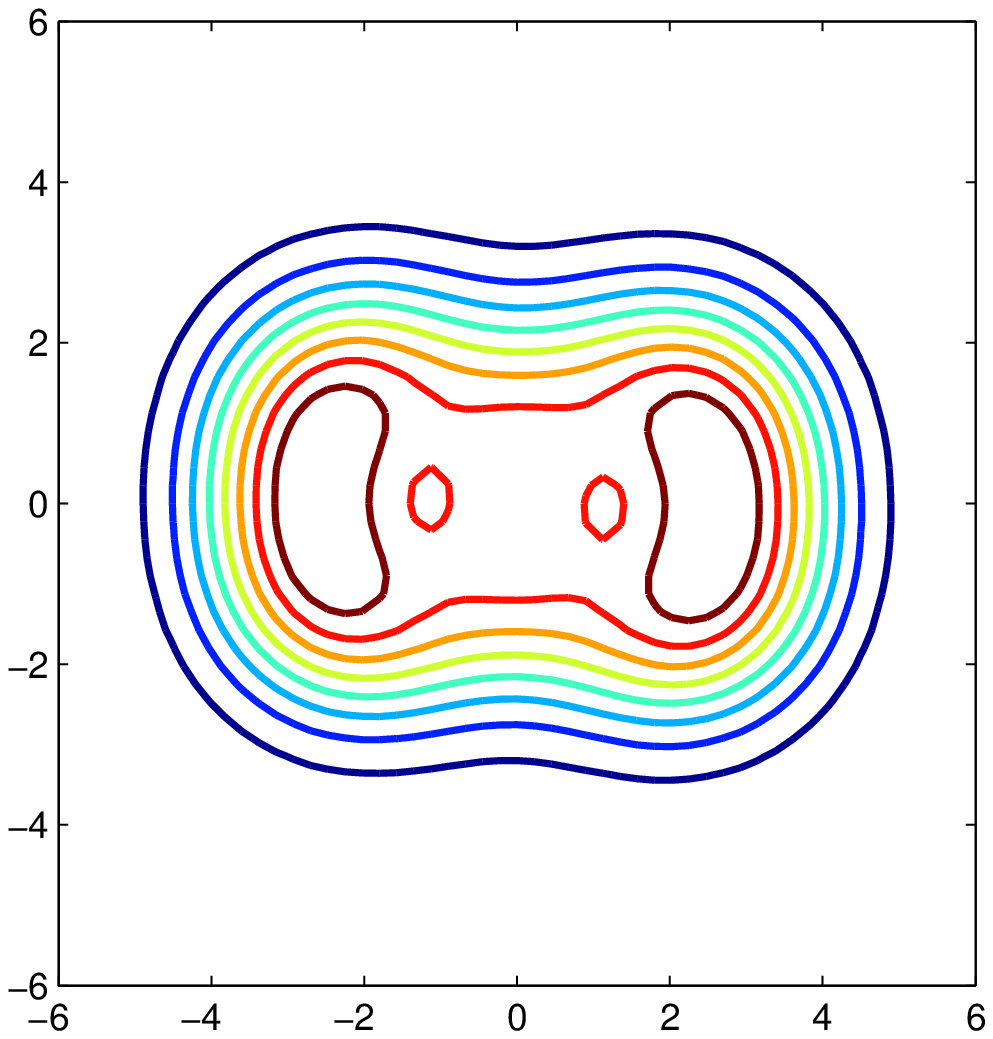}\quad\includegraphics[width=0.45\textwidth]{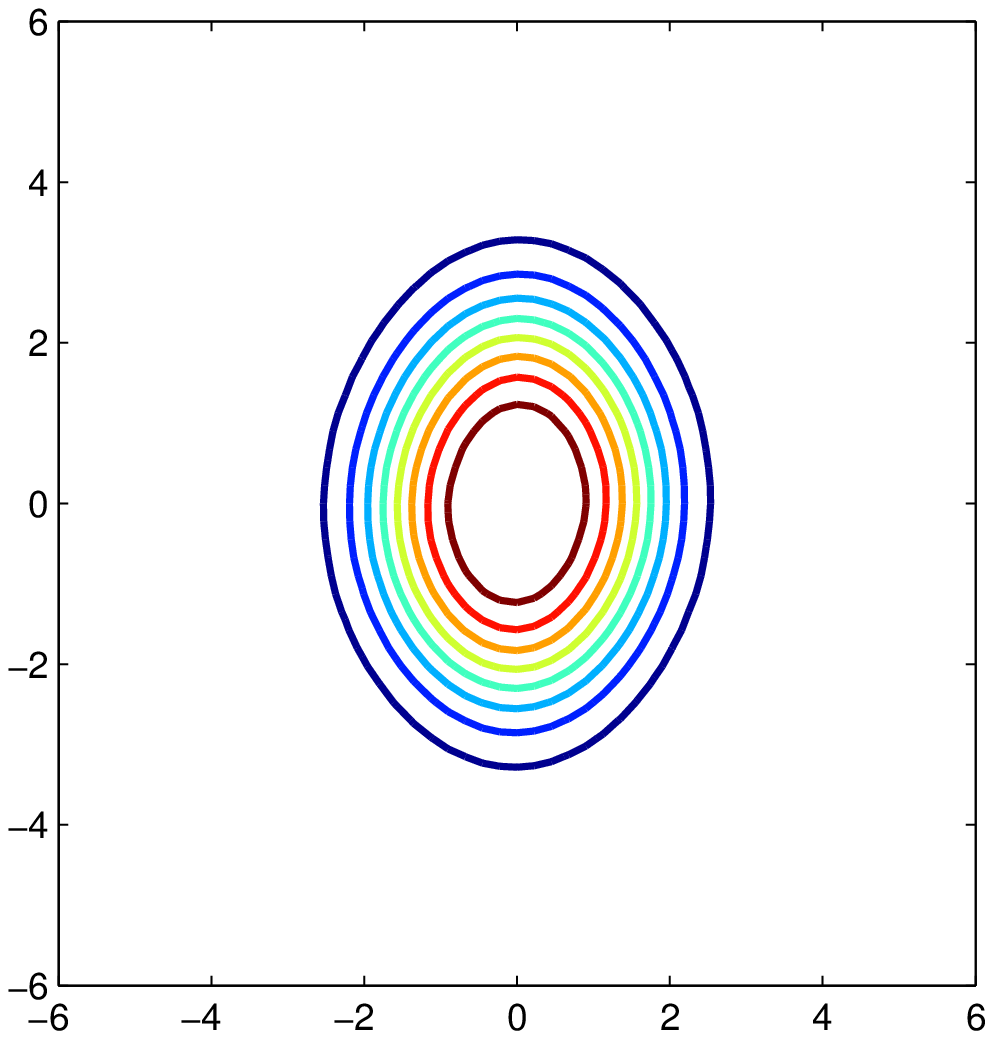}}
\caption{(Color online) Contour plots of the neutron density distributions of $^{24}$Ne in a plane parallel (left panel) and perpendicular (right panel) to the elongation axis. The contour lines represent 8 densities ranging in equal step-sizes between zero and maximal density.}
\label{fig:Ne24el}
\end{center}
\end{figure}

To visualize more details about the symmetry of these density distributions, Fig.~\ref{fig:Ne24el} displays the density distribution in a plane, which contains the axes of prolate elongation (left panel) and on perpendicular to the elongation axis. These plots demonstrate that the resulting density distributions are indeed triaxial and
could not be described assuming a cylindrical symmetry, as has been tried in the work of Ref.~\cite{ebran:2011}. The same can be observed for the other isotopes.

What is the gain in energy, which is related to the deformation of the nuclei? To answer this question, we have supplemented the calculations in a cartesian box, which we discussed so far with corresponding calculations in a spherical cavity (see \cite{vandalen:2009b} and discussion around eq.(\ref{eq:sbasis}) above) assuming spherical symmetry and, if necessary, a partial occupation of the degenerate states close to the Fermi energy.

\begin{figure}[!h]
\begin{center}
\includegraphics[width=0.5\textwidth] {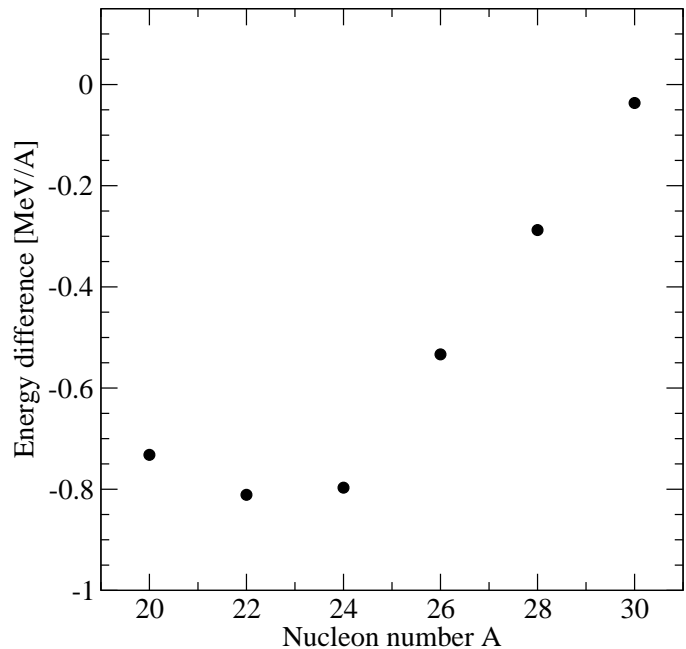}
\caption{(Color online) The energy difference between calculations allowing for tri-axial structures and calculations allowing only spherical structures. The binding energy results considered are those without cm correction in both cases.}
\label{fig:difEbind}
\end{center}
\end{figure}

The tri-axial calculations predict stable isotopes up to $^{32}$Ne. The HF calculations allowing only for spherical structures gives stable isotopes up to $^{30}$Ne. Therefore, the differences in the binding energies between tri-axial calculations and spherical calculations are plotted in Fig.~\ref{fig:difEbind} only for isotopes 
up to $^{30}$Ne. The total energy gain allowing for tri-axial deformation is largest in the case of $^{24}$Ne (0.8 MeV per nucleon, which corresponds to 19.2 MeV in total). The energy gain is reduced for the heavier isotopes and is close to zero in the case of $^{30}$Ne, in which all neutron states of a spherical $sd$-shell can be filled (see
also discussion of the radii of the neutron distributions above). 

\begin{figure}[!h]
\begin{center}
\hbox{\includegraphics[width=0.45\textwidth]{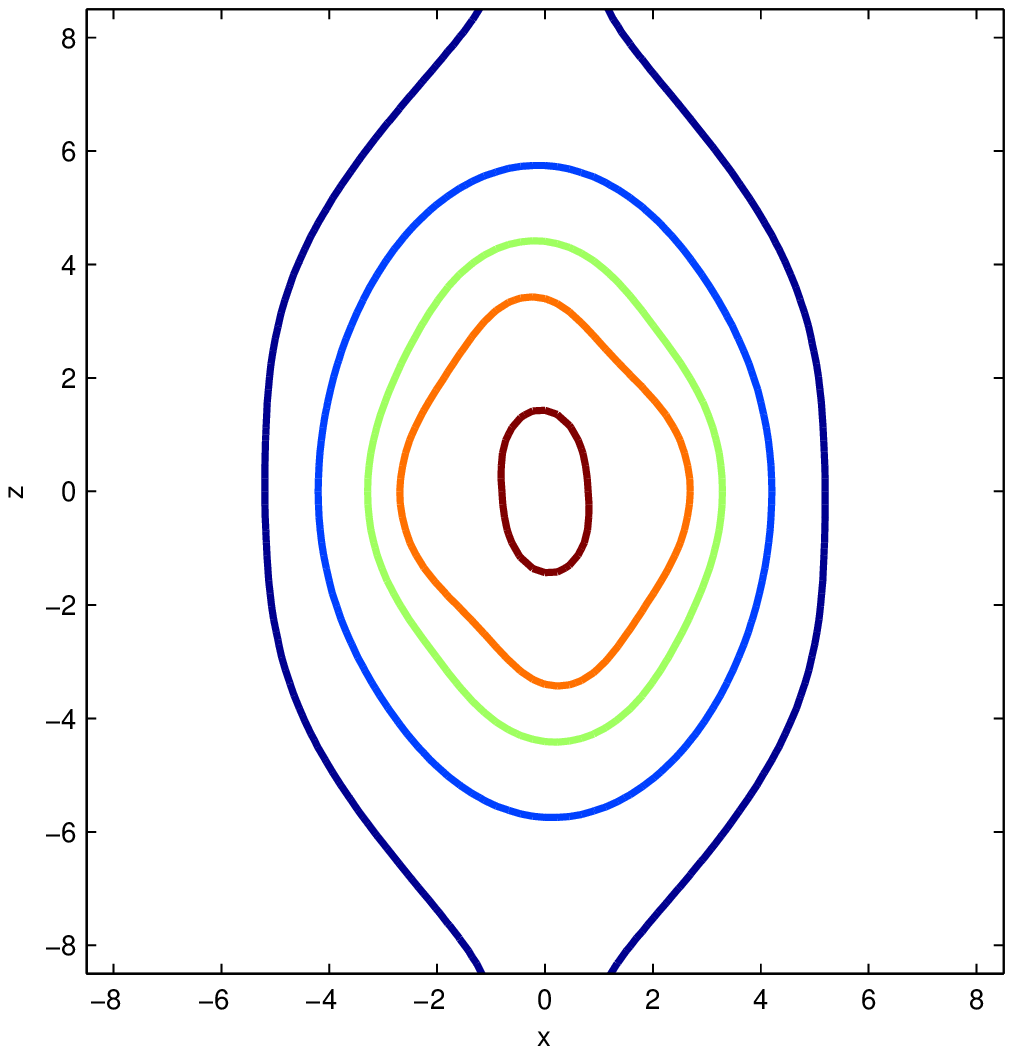},\includegraphics[width=0.45\textwidth]{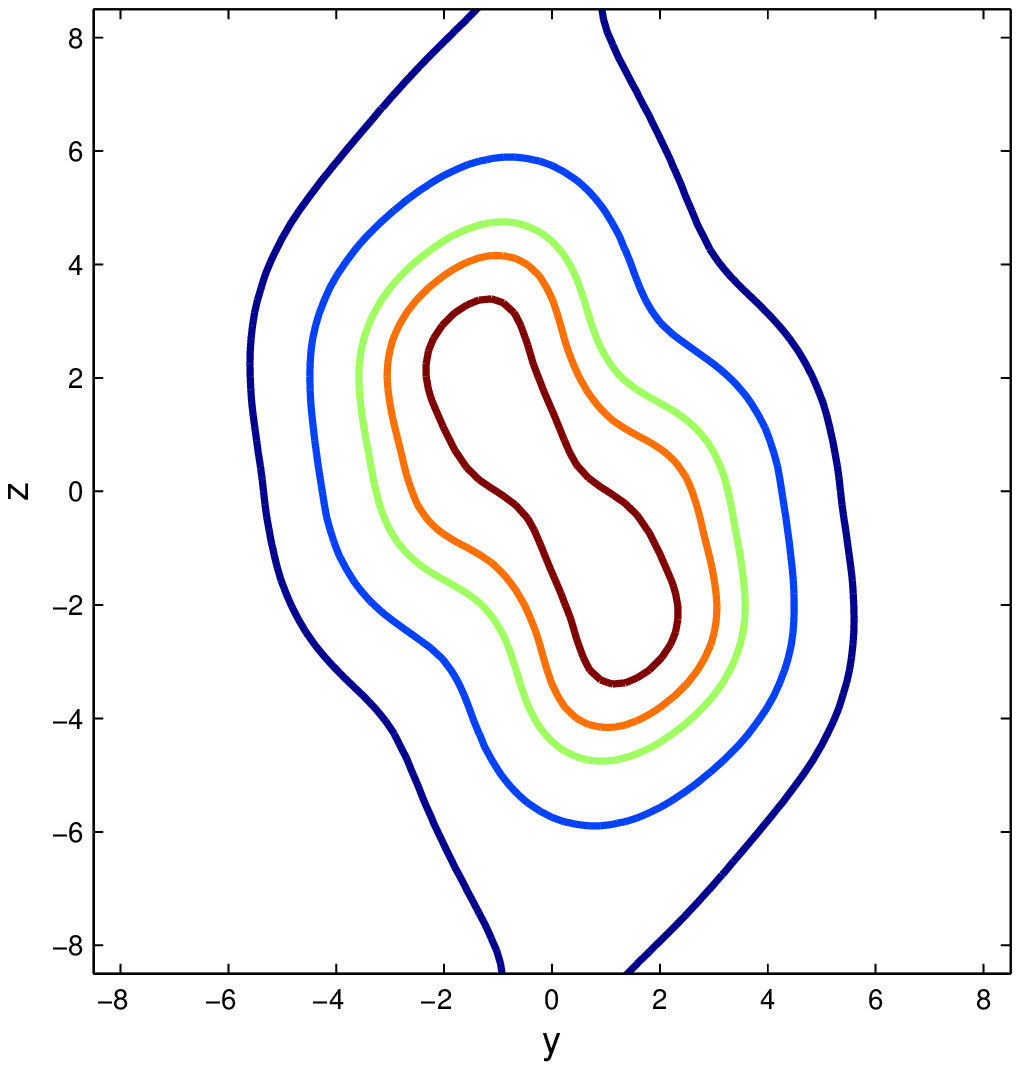}}
\caption{(Color online) Contour plots of the neutron density distributions of a quasi-nuclear structure with 10 protons and 32 neutrons in the xz plane (left panel) and yz plane (right panel) of the box considered in the calculation. The contour lines represent the densities 0.005, 0.02 0.045, 0.07, and 0.09 nucleons fm$^{-3}$, respectively.}
\label{fig:Ne42}
\end{center}
\end{figure}

In the calculation it is easy to go beyond the description of stable neon isotopes to the situation of nuclear structures with unbound neutrons. This should have some relevance for the transition from discrete nuclei to homogeneous matter which is
supposed to occur, e.g. in the crust of neutron stars. 
As an example of a system with unbound neutrons we consider a system with 10 protons and 32 neutrons. Results for the density distributions for the neutrons are plotted in Fig.~\ref{fig:Ne42}.
In this system, the density profile in z direction deviates clearly from that in the x and y direction. 
One has a quasi-nucleus at the
center of the cubic box with a region of a low-density neutron gas extending in the $z$ direction to  the boundary of the box. Since the box displayed must be considered as one member of a lattice of boxes with periodic boundary conditions, this region of a low-density neutron gas continues to the boxes neighbored in $z$-direction, forming a rod of neutrons with quasi-nuclear structures located along this rod.

This observation is of interest in the sense that it demonstrates that the gas of evaporated neutrons may show structures like the one discussed here. However, the present calculation is not realistic in the sense that the cartesian boxes considered in our calculation are to small to simulate the structure of quasi-nuclei embedded in a gas of quasi-free neutrons in the crust of neutron stars. Furthermore, to apply such calculations to the neutron star crust
the contribution of leptons has to be included and the $\beta$-equilibrium condition has to be imposed.

\section{Summary and conclusion}
\label{sec:SC}

The aim of this study is to establish a general tool, which allows a Hartree-Fock (HF) calculation of weakly bound open shell nuclei based on a general realistic NN interaction. For that purpose the HF equations have been solved in basis of plane wave states with periodic boundary conditions in a cartesian box. Stable results are obtained for light nuclei employing a box with a box size of around 20 fm and basis size of around 1500 states for protons and neutrons, respectively. This allows the description of structures with general symmetry. 
The variational procedure is restricted to single-particle states with good parity. This stabilizes the position of the center of mass of the nucleus to the center of the box.

The model used for the NN interaction is a renormalized low momentum potential, $V_{lowk}$, derived from a realistic NN interaction which fits the NN scattering data. This $V_{lowk}$ interaction has been supplemented with a contact interaction with three parameters adjusted to reproduce the empirical saturation point of symmetric nuclear matter and the symmetry energy in nuclear matter.

Calculations have been performed for Ne isotopes with even number of neutrons, ranging from nucleon number
$A$=18 to $A$=34. Results for the bulk properties of these nuclei are very encouraging. The total energy is reproduced with an accuracy of better than 2 percent and also the radii of the charge distributions of the stable isotopes are described with high accuracy. The calculations reproduce the position of the neutron drip-line for the Ne isotopes. 

The deformation degrees of freedom are very important providing an extra binding energy up to almost 20 MeV as
compared to a calculation restricted to spherical symmetry with partial occupation probabilities for degenerate states at the Fermi energy. The shapes of the nuclear density distributions are tri-axial with a
prolate elongation axis. 

In a next step one may perform Hartree-Fock-Bogoliubov calculations, imposing the time-reversal symmetry in the HF variation and using the realistic $V_{lowk}$ for a consistent description of pairing properties. This would also enable us to investigate the role of pairing in weakly bound nuclei. Also the study of isotope chains different from Ne may be of interest.

\section{Acknowledgments}
This work has been supported by a grant (Mu 705/7-1) of the Deutsche
Forschungsgemeinschaft DFG.


\begin{thebibliography}{99}

%\bibitem{Oyamatsu:2007}{K. Oyamatsu and K. Iida, Phys. Rev. {\bf C75},  015801 (2007).}

%\bibitem{Montani:2004}{F. Montani, C. May, and H. M\"{u}ther, Phys. Rev. C {\bf 69}, 065801 (2004).}

%\bibitem{Goegelein:2007}{P. G\"{o}gelein and H. M\"{u}ther, Phys. Rev. C{\bf 77},
%024312 (2008).}

%\bibitem{goegeleina:2008}{P. G\"ogelein, E. N. E. van Dalen, C. Fuchs, and H.
%M\"uther, Phys. Rev. C \textbf{77}, 025802 (2008).}

\bibitem{caurier:2008} E. Caurier, J. Men$\acute{e}$ndez, F. Nowacki, and A. Poves, Phys. Rev.  Lett. \textbf{100},
052503 (2008).

\bibitem{bender:2003} M. Bender, P.-H. Heenen, and P.-G. Reinhard, Rev. Mod. Phys. \textbf{75}, 121 (2003).

\bibitem{niksic:2011} T. Niksic, D. Vretenar, and P. Ring, Prog. Part. Nucl. Phys. \textbf{66}, 519 (2011).

\bibitem{arg1} R.B. Wiringa, V.G.J. Stoks, and R. Schiavilla, Phys. Rev. C
\textbf{51}, 38 (1995).

\bibitem{mach1} R. Machleidt, F. Sammarruca, and Y. Song, Phys. Rev. C\textbf{ 53},
R1483 (1996).

\bibitem{chirar} R. Machleidt and D.R. Entem, Phys. Rep. \textbf{503}, 1 (2011).

\bibitem{chira1} E. Epelbaum, H.-W. Hammer, and U.-G. Mei{\ss}ner, Rev. Mod. Phys. \textbf{81}, 1773 (2009).

\bibitem{chira2} N. Kaiser, S. Fritsch, and W. Weise, Nucl. Phys. \textbf{A 697}, 255 (2002).

\bibitem{polls00}{ H. M\"uther and A. Polls, Prog. Part. Nucl. Phys. {\bf 45},
	243 (2000).}

\bibitem{vandalen:2009b} E. N. E. van Dalen, P. G\"ogelein, and
  H. M\"uther, Phys. Rev. C \textbf{80}, 044312  (2009).

\bibitem{bogner:2001} S. K. Bogner, T.T.S. Kuo, and L. Coraggio, Nucl. Phys.
\textbf{A684}, 432c (2001).

\bibitem{bogner:2005}{S.K. Bogner, A. Schwenk, R.J. Furnstahl, and A. Nogga, Nucl.
Phys. \textbf{A763}, 59 (2005).}

\bibitem{bogner:2007}{S. K. Bogner, R. J. Furnstahl, S. Ramanan, and A. Schwenk,
Nucl. Phys. \textbf{A784}, 79 (2007).}

\bibitem{bozek:2006}{P. Bo\.{z}ek, D.J. Dean, and H. M\"uther, Phys. Rev. C
\textbf{74}, 014303 (2006).}

\bibitem{vandalen:2010} E. N. E. van Dalen and H. M¨uther, IJMPE \textbf{19}, 2077 (2010).

\bibitem{kuckei:2003} J. Kuckei, F. Montani, H. M\"uther, and A. Sedrakian, Nucl.
Phys. \textbf {A 723}, 32 (2003).

\bibitem{sk1} T.H.R. Skyrme, Nucl. Phys. \textbf{9}, 615 (1959).

\bibitem{bv81} P. Bonche and D. Vautherin, Nucl Phys. \textbf{A 372}, 496 (1981).

\bibitem{coraggio:2005}{L. Coraggio, A. Covello, A. Gargano, N. Itaco, T. T. S.
Kuo, and R. Machleidt, Phys. Rev. C \textbf{71}, 014307 (2005).}

\bibitem{coraggio:2006}{L. Coraggio, A. Covello, A. Gargano, N. Itaco, and T. T. S.
Kuo, Phys. Rev. C \textbf{75}, 057303 (2007); Phys. Rev. C \textbf{73}, 014304
(2006).}

\bibitem{martini:2011} M. Martini, S. P\'eru, and M. Dupuis, Phys. Rev. C \textbf{83}, 034309 (2011).

\bibitem{ebran:2011} J.-P. Ebran, E. Khan, D. Pe\~na Arteaga, and D. Vretenar, Phys. Rev. C \textbf{83}, 064323 (2011).

\bibitem{suzuki:1982}{K. Suzuki, Prog. Theoret. Phys. \textbf{68}, 246 (1986).}

\bibitem{fuji:2004} S. Fujii, R. Okamoto, and K. Suzuki, Phys. Rev. C \textbf{69},
034328 (2004).
\bibitem{pgreinh} P.-G. Reinhard in {\it Computational Nuclear Physics I}, Eds.: K. Langanke, J.A. Maruhn, and S.E. Koonin, Springer Verlag, 28 (1991).

\bibitem{Bonche:2005} P. Bonche, H. Flocard, P.-H. Heenen, Comp. Phys. Comm. {\textbf 171}, 49 (2005).

\bibitem{Ryssens:2014} W. Ryssens, V. Hellemans, B. Bender, and P.-H. Heenen, arXiv:1405.1897v1 (2014).

\bibitem{audi:1993} G. Audi and W. H. Wapstra, Nucl. Phys. \textbf{A565}, 1 (1993).

\bibitem{marinova} K. Marinova, W. Geithner, M. Kowalska, K. Blaum, S. Kappertz, M. Keim, S. Kloos, G. Kotrotsios, P. Lievens, R. Neugart, H. Simon, and S. Wilbert, Phys. Rev. C \textbf{84}, 034313 (2011).
 
%************************************

%\bibitem{Pieper:2006} S.C. Pieper, R.B. Wiringa, and J. Carlson, Phys. Rev. C
%\textbf{70}, 054325 (2006).

%\bibitem{vandalen:2009a} E. N. E. van Dalen and H. M\"uther, Phys. Rev C \textbf{80}, 037303 (2009).

%\bibitem{gogelein:2009} P. G\"ogelein, E.N.E. van Dalen, Kh. Gad, Kh.S.A. Hassaneen, and H. M\"uther,
%Phys. Rev C \textbf{79}, 024308 (2009).




%\bibitem{fuji:2004} S. Fujii, R. Okamoto, and K. Suzuki, Phys. Rev. C \textbf{69},
%034328 (2004).

%\bibitem{audi:1993} G. Audi and W. H. Wapstra, Nucl. Phys. \textbf{A565}, 1 (1993).





%\bibitem{Kung:1979} {C.L. Kung, T.T.S. Kuo, and K.F. Ratcliff, Phys. Rev. C
%\textbf{19}, 1063 (1979).}

%\bibitem{bonatsos:1989} {D. Bonatsos and H. M\"uther, Nucl. Phys. \textbf{A496},
%23 (1989).}

%\bibitem{vandalen:2004b} E. N. E. van Dalen, C. Fuchs, and Amand Faessler, Nucl.
%Phys. \textbf{A744}, 227 (2004).

%\bibitem{vandalen:2007}{E.N.E. van Dalen, C. Fuchs, and  A. Faessler, Eur. Phys. J.
%A \textbf{31}, 29 (2007).}

%\bibitem{lejeune:2000} A. Lejeune, U. Lombardo, and W. Zuo,  Phys.Lett. B
%\textbf{477}, 45 (2000).

%\bibitem{coester:1970} F. Coester, S. Cohen, B.D. Day, and C.M. Vincent, Phys. Rev. C
%\textbf{1}, 769 (1970).

%\bibitem{brock:1984} R. Brockmann and R. Machleidt, Phys. Lett. B {\bf
%149}, 283 (1984).

%\bibitem{muemach:1990} H. M\"uther, R. Machleidt, and R. Brockmann,
%Phys. Rev. C{\bf 42}, 1981 (1990).

%\bibitem{stoicea:2004} G. Stoicea {\it et al.} [FOPI Coll.], Phys. Rev. Lett. {\bf 92},
%072303 (2004).

%\bibitem{sturm:2001} C. Sturm {\it et al.} [KaoS Coll.], Phys. Rev. Lett. {\bf 86},
%39 (2001); C. Fuchs, A. Faessler, E. Zabrodin, Y.M. Zheng, Phys. Rev. Lett. {\bf
%86}, 1974 (2001); C. Fuchs, Prog. Part. Nucl. Phys. {\bf 56}, 1 (2006).



%\bibitem{hofmann:2001}{F. Hofmann, C.M. Keil, and H. Lenske, Phys. Rev. C
%\textbf{64}, 034314  (2001).}

%\bibitem{coraggio:2003}{L. Coraggio, N. Itaco, A. Covello, A. Gargano, and T. T.
%S. Kuo, Phys. Rev. C \textbf{68}, 034320 (2003).}













\end{thebibliography}
\end{document}